\RequirePackage{fixltx2e}
\documentclass[nofootinbib,reprint,showpacs,
  prc,aps,10pt,
  altaffilletter,floatfix]{revtex4-1}
\usepackage{graphicx}
\usepackage{hyperref}
\usepackage{amsmath}
\usepackage{amssymb}
\usepackage[amsmath,thmmarks]{ntheorem}
\usepackage{bm}
\usepackage{cleveref}
\usepackage[caption=false]{subfig}
\usepackage{array}
\usepackage{ifthen}
\usepackage{booktabs}
\usepackage{textgreek}
\usepackage{multirow}
\usepackage{enumitem}
\usepackage{dcolumn}
\usepackage{xspace}

\usepackage{color}

\makeatletter
\g@addto@macro\bfseries{\boldmath}
\makeatother

\begin{document}

\newcommand{\comment}[1]{} 

\newcommand{\nue}       {$\nu_{e}$\xspace}
\newcommand{\numu}      {$\nu_{\mu}$\xspace}
\newcommand{\nutau}     {$\nu_{\tau}$\xspace}
\newcommand{\nubar}     {$\overline{\nu}$\xspace}
\newcommand{\nuebar}    {$\overline{\nu}_{e}$\xspace}
\newcommand{\numubar}   {$\overline{\nu}_{\mu}$\xspace}
\newcommand{\nutaubar}  {$\overline{\nu}_{\tau}$\xspace}
\newcommand{\znbb}      {$0\nu\beta\beta$\xspace}
\newcommand{\tnbb}      {$2\nu\beta\beta$\xspace}
\newcommand{\otsx}      {$^{136}\mathrm{Xe}$\xspace}
\newcommand{\dd}        {\mathrm{d}}
\newcommand{\rpm}{\raisebox{.2ex}{$\scriptstyle\pm$}}
\newcommand{\red}       {\textcolor{red}}
\newcommand{\mcl}       {$M_{\mathrm{cl}}$\xspace}
\newcommand{\mmmcl}       {M_{\mathrm{cl}}}

\title{Neutron inelastic scattering measurements on \otsx at $E_{n} =$ 0.7~to~100~MeV}

\author{S. J. Daugherty}\email{seajdaug@indiana.edu}
\author{J. B. Albert}
\author{L. J. Kaufman}
\altaffiliation{Also at SLAC National Accelerator Laboratory}
\affiliation{Physics Department and CEEM, Indiana University, Bloomington, 
Indiana 47405, USA}

\author{M. Devlin}
\author{N. Fotiades}
\author{R. O. Nelson}
\affiliation{Los Alamos National Laboratory, Los Alamos, New Mexico 87545, USA}

\author{M. Krti\v{c}ka}
\affiliation{Faculty of Mathematics and Physics, Charles University, CZ-180 00 Prague, Czech Republic}

\date{\today}

\begin{abstract}

Experiments searching for neutrinoless double beta decay (\znbb) require
precise energy calibration and extremely low backgrounds.  One of the most popular
isotopes for \znbb experiments is \otsx. In support of these
experiments, the neutron inelastic
scattering properties of this isotope have been measured at the GErmanium Array for
Neutron Induced Excitations (GEANIE) at the Los Alamos Neutron Science Center.
Time-of-flight techniques are utilized with high-purity germanium detectors 
to search for inelastic scattering 
$\gamma$ rays for neutron energies between 0.7 and 100~MeV.
Limits are set on production of yet-unobserved $\gamma$ rays in the energy range
critical for \znbb studies, and measurements are made of multiple $\gamma$-ray
production cross sections.  In particular, we have measured the production of
the 1313~keV $\gamma$ ray which comes from the transition of the first-excited
to ground state of \otsx.  This neutron-induced $\gamma$ line may be
useful for a novel energy calibration technique, described in this paper.

\end{abstract}

\maketitle


\section{\label{sec:intro}Introduction}

Neutrinoless double beta decay (\znbb) is a hypothetical 
lepton-number-violating decay mode of great interest to nuclear and particle physics.
Its observation would confirm that neutrinos are Majorana particles, meaning
there is no distinction between neutrinos and antineutrinos.  Measurement of
this process could, in the Majorana neutrino case, also be used to infer the
absolute mass of the neutrino.  Given the interest in these neutrino properties,
several experimental collaborations are running
or developing experiments to search for \znbb.

One of the most popular and successful isotopes for this search is 
\otsx~\cite{Kharusi:2018eqi, Brunner:2017iql, PhysRevC.97.065503, PhysRevLett.117.082503, Martin-Albo:2015rhw, Chen2017}.
This isotope has several advantages, including the large 
$Q$ value (2457.83~keV~\cite{Redshaw:2007un}),
ease of enrichment and purification, and physical characteristics allowing for scaling 
to large monolithic detectors.
While the signal for \znbb is a spatially-compact mono-energetic peak at the $Q$ value, 
signals detected at other energies and with spatially-separated (multi-site) energy
deposits can be used to identify and constrain backgrounds.
Proposed next-generation experiments (such as nEXO~\cite{Kharusi:2018eqi}) will
have reduced radiogenic backgrounds
(such as the 2448~keV line from $^{214}$Bi decay) due to purification,
high-radiopurity shielding, topological discrimination, and other techniques.
With these backgrounds mitigated, less common backgrounds will become more prominent.
Current and future experiments using \otsx (including nEXO, NEXT~\cite{Martin-Albo:2015rhw}, 
PandaX-III~\cite{Chen2017}, and KamLAND-Zen~\cite{PhysRevLett.117.082503}) will benefit
from having the best possible understanding of all potential backgrounds.

As such, we have
measured the production of $\gamma$ rays in \otsx from fast neutron interactions.
This complements previous work~\cite{Albert:2016vmb} studying $\gamma$ rays from
neutron capture on \otsx.  While neutron-induced $\gamma$ rays
are not expected to be a significant background in future deep underground detectors, a detailed
understanding of them may help identify otherwise mysterious signals.  Cross sections
as a function of neutron energy have been measured for several of the major
$\gamma$ ray energies, and limits have been set on the potential cross sections
of unknown $\gamma$ lines which might mimic a \znbb signal.
The measured cross sections for $\gamma$ lines produced in fast neutron-induced reactions 
may be used to refine future nuclear evaluations and improve simulations of the reactions.

This work also suggests a novel method for detector energy calibration.  The $1313$~keV 
level of \otsx is the first-excited state, and can be reached through neutron
inelastic scattering $(n,n')$.  Thus, fast neutrons could be produced and sent
into the detector to excite \otsx nuclei to this state, causing them to emit
this mono-energetic $\gamma$ line, offering a calibration signal.

\section{\label{sec:experiment}Experimental Setup}

Data were collected at the Los Alamos Neutron Science Center (LANSCE) Weapons Neutron
Research (WNR) facility~\cite{doi:10.13182/NSE90-A27471}.  There, an 800-MeV
proton beam incident on a natural tungsten target produces neutrons in a wide
energy range.  The proton beam is delivered in short pulses spaced 1.8~$\mu$s
apart (micropulses), in groups lasting for 625~$\mu$s (macropulses).  There are
100 macropulses per second.

Our sample of 99.925\% enriched \otsx was contained in a thin cylindrical 
aluminum vessel with thin kapton windows for neutrons to pass through.  The target vessel
window diameter of 3.4~cm is larger than the measured beam diameter, so the precise beam profile
is not relevant to the cross-section measurement.  The vessel was pressurized to
near 2700~torr absolute, with pressure monitored 
in real time using a capacitance manometer to account for changes due to
temperature and a very slow leak.  The xenon volume
is 7.2~cm long and 3.4~cm diameter, located on flight path 60R of the WNR neutron beam, centered 20.34~m downstream
of the proton target, in the center of the 
GErmanium Array for Neutron Induced Excitations (GEANIE) 
~spectrometer~\cite{doi:10.1080/10506899709410550}.

\begin{figure}[htb]
    \begin{center}
    \includegraphics[clip, keepaspectratio=true, width=\columnwidth, trim={0.18cm 0.03cm 0.2cm 0.5cm}]{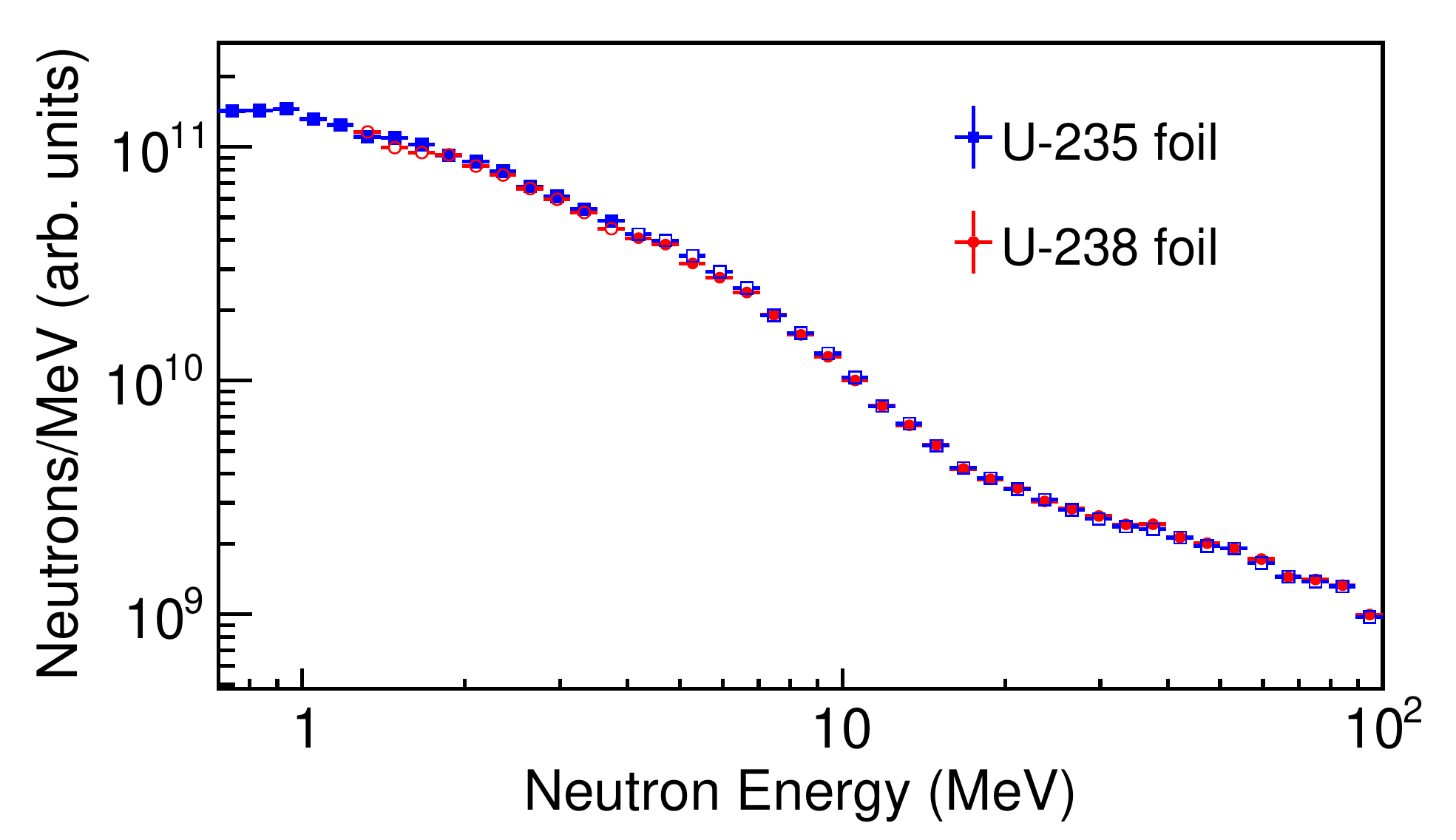}
    \end{center}
    \caption[flux]
    {Measured neutron flux spectrum at GEANIE.  The relative flux magnitudes for the
    xenon and iron calibration datasets were determined exclusively using the $^{235}\mathrm{U}$ 
    fission foil data, as the $^{238}\mathrm{U}$ efficiency was less stable.  This figure is
    not normalized to any specific set of beam time.  The flux shape was determined
    by the $^{235}\mathrm{U}$ ($^{238}\mathrm{U}$) data for $E_{n} \le 4$~MeV ($E_{n} > 4$~MeV).  
    Open markers are used here to demonstrate flux shape agreement between
    the two measurements.
    (color online)}
     \label{fig:fluxplot}
\end{figure}

GEANIE features an array of 20 Compton-suppressed high purity germanium (HPGe) detectors to detect
neutron-induced $\gamma$ rays.  Neutron energy is determined via time of flight.  
Neutron flux is measured by a fission chamber in the beam line upstream from
the target with $^{235}\mathrm{U}$ and $^{238}\mathrm{U}$
foils~\cite{WENDER1993226}.  Figure~\ref{fig:fluxplot} shows the measured
neutron flux as a function of neutron energy.  More details on GEANIE, the WNR beam,
and the procedures for cross-section measurement can be found in other
papers~\cite{Guiseppe:2008qj, Fotiades:2004yi, MacMullin2013, PhysRevC.85.064614}.

\section{\label{sec:data_ana}Data Analysis}

\subsection{\label{sec:absxs}Absolute Cross Sections}

To minimize uncertainties due to absolute detector efficiency and neutron flux,
we normalized our xenon cross-section measurements to a reference $^{56}\mathrm{Fe}$ cross section.
To do so, we added
50~$\mu$m thick natural iron foils to the ends of the target vessel for certain runs.  
Our reference cross section was $1.5\pm0.1$~b for
production of the 847~keV $\gamma$ in $^{56}$Fe at
$E_{n} = 6.2$~MeV, taken from a measurement by Beyer~$et~al.$~\cite{Beyer:2014ota} at the
nELBE photoneutron source.
This $\gamma$ line is the transition from the first-excited to ground state of 
$^{56}\mathrm{Fe}$.
We corrected this cross section for small angular variations using angular coefficients from
another measurement at nELBE, by Dietz~$et~al.$~\cite{Dietz:56festuff}.  After
corrections for detector live-time, this line provided a reliable calibration for 
measurement of the absolute
cross section.  Our vessel was filled with
nitrogen during most of the iron foil calibration runs to keep neutron scattering rates similar
to xenon while avoiding any possible interference due to
the 847~keV $\gamma$-rays emitted from $^{134}$Xe produced via $(n,3n)$ interactions.
We also took some iron foil data with the vessel filled with xenon to confirm
the validity of our analysis.  Lastly, runs with the target vessel filled with nitrogen
(both with and without iron foils) were used to identify backgrounds not due
to neutron interactions on xenon.

The relative detector efficiencies at different $\gamma$ energies were determined
using a $^{152}$Eu source placed at the center of the detector array.  This
provided $\gamma$ rays with known relative intensities between 444 and 1299~keV.
We determined an efficiency curve by fitting the data with~\cite{RadWare}
\begin{equation}
\log(\epsilon_{\gamma}) = a + b \log(E_{\gamma}) + c \log^{2}(E_{\gamma}).
\end{equation}
Successful extrapolation of this efficiency curve to higher $\gamma$-ray energies has been
demonstrated in previous GEANIE measurements~\cite{MacMullin2013, Fotiades:2013ama}.
The known $^{152}$Eu lines, along with some higher
energy lines from $(n, n')$ interactions on \otsx, were also used to calibrate
the conversion from ADC counts to measured $\gamma$-ray energy.

The GEANIE spectrometer includes both coaxial and planar HPGe detectors, though we
utilized only the coaxial ones. The planar detectors can only measure
$\gamma$ rays up to 1~MeV, well below the energies of interest to \znbb experiments.
Some of the coaxial detectors showed poor performance in neutron or
$\gamma$ energy reconstruction, possibly due to neutron damage, and were excluded
from the analysis, leaving five detectors with usable data.

\begin{figure}[htb]
    \begin{center}
    \includegraphics[clip, keepaspectratio=true, width=\columnwidth, trim={0.05cm 0.0cm 1.8cm 0.0cm}]{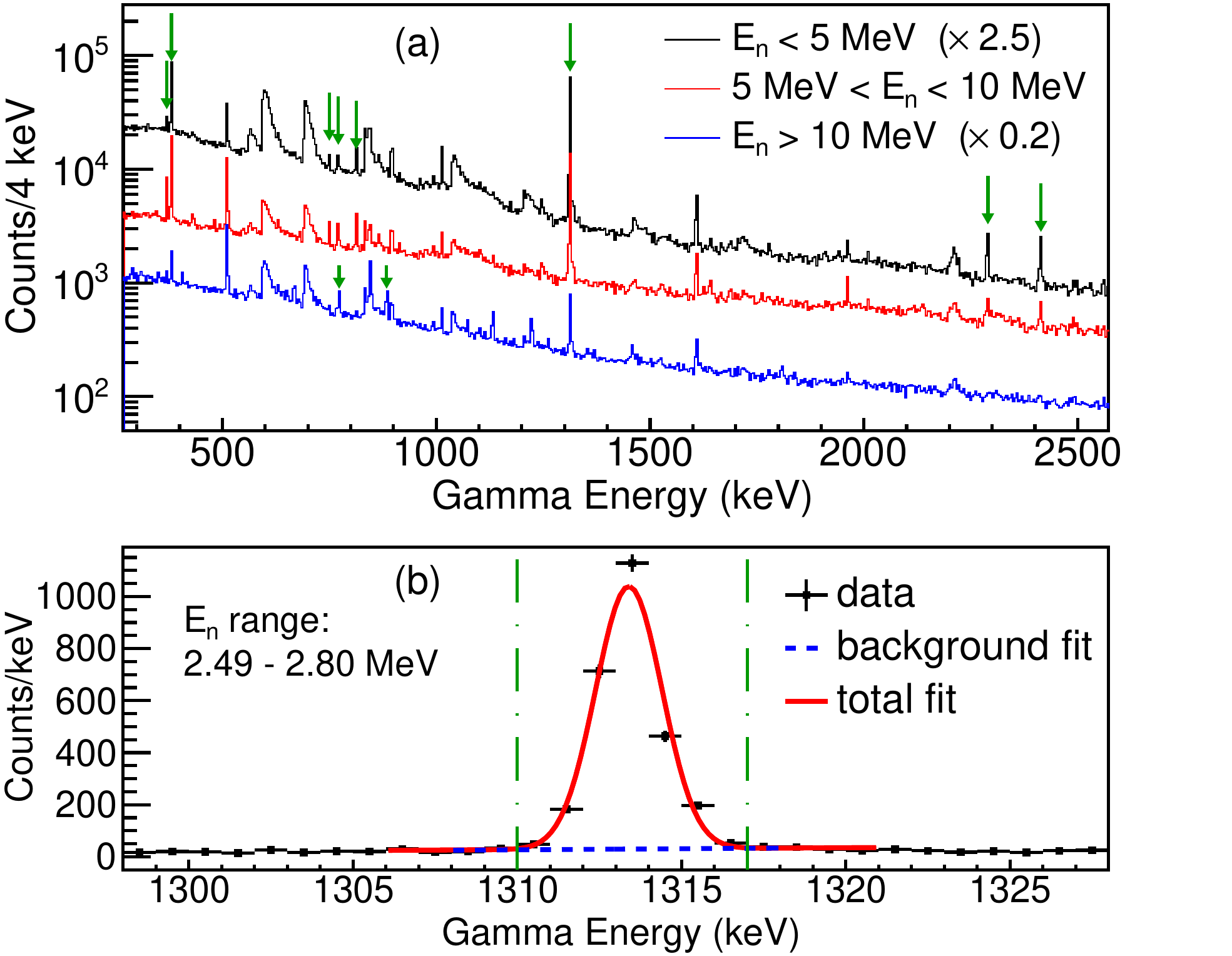}
    \end{center}
    \caption[Detector P examples]
    {(a) Spectra of $\gamma$ rays detected by one of the HPGe detectors.
    Three different neutron energy ranges are shown: $E_{n} < 5$~MeV (top),
    5~MeV$ < E_{n} < 10$~MeV (middle) and $E_{n} > 10$~MeV (bottom).  
    The top (bottom) curve has been scaled by
    a factor of 2.5 (0.2) to make it easier to distinguish the spectra.  Downward
    pointing arrows indicate the peaks for which we have determined cross sections
    in this work.  Other peaks visible in this figure are also present in data taken
    without xenon in the target vessel, and thus are considered to be background lines.
    (b) A fit to the 1313~keV line, for a single bin in $E_{n}$.  The
    vertical axis gives the detected counts in a single detector.  
    The solid and dashed lines show the 
    total fit and background contribution, respectively, and the vertical broken lines indicate the boundaries of
    the integration window for the peak.
    (color online)}
     \label{fig:detPexamples}
\end{figure}

A simple Gaussian-plus-linear fit is performed for each bin in $E_{n}$ for each $\gamma$
peak to define the integration range in each of the HPGe detectors used.  
The $\gamma$ yield is then determined by summing counts in the peak
range of $\pm 3.5 \sigma$ and subtracting the background (estimated as the linear part of the fit).  
Figure~\ref{fig:detPexamples} shows example $\gamma$ spectra and an example fit.  
Live-time fractions for the fission chambers and HPGe detectors are
determined for each run by comparing ADC triggers with dead-time free scaler counts.
The measured yields, live-times, efficiencies, and $\gamma$-ray attenuation 
corrections are combined similarly to the method described in Ref.~\cite{Guiseppe:2008qj} to determine
partial gamma-ray cross sections.  Our analysis differs in that we normalize to an iron cross section,
rather than lead, and that we neglect internal conversion coefficients, looking only at $\gamma$ production.
The influence of internal conversion is very weak 
(much smaller than measurement uncertainties) for all measured $\gamma$-rays~\cite{KIBEDI2008202}.

Our analysis also differs
in the method of dealing with anisotropic $\gamma$ emission.  
The neutron-induced $\gamma$ emission will not be isotropic,
but rather will have some angular distribution which may be, for known spin states, 
estimated.  The estimation of these angular distributions is non-trivial,
and we chose instead to minimize the angular effects where possible, and accept 
a larger uncertainty elsewhere.  We model the angular correction, as is typical,
with
\begin{equation}
C_{\gamma}(E_{n}, \theta) = \frac{1}{W(E_{n}, \theta)};
\end{equation}
\begin{equation}
W(E_{n}, \theta) = 1 + A_{2} P_{2}\left(\cos\theta\right) + A_{4} P_{4} \left(\cos\theta\right),
\end{equation}
where $A_{k}$ are coefficients and $P_{k}$ are the Legendre polynomials~\cite{Dietz:56festuff}.  
With cross-section 
measurements at three distinct $\left|\cos\theta \right|$ values, 
it is possible to add the three measurements linearly with coefficients $\beta_{i}$
for different detectors $i$
such that the deviations from $C_{\gamma}(E_{n}, \theta) = 1$ will cancel out,
regardless of the $A_{k}$ coefficients.  This calculation uses the approximation $1/(1+x) \approx 1 - x$ for 
small $x = A_{2}P_{2} + A_{4} P_{4}$, and is accurate to 
order $x^{2}$.  Among our five detectors, we have three
detectors at nearly the same value of $\left|\cos\theta \right|$, and two others
at sufficiently different values.  Unfortunately, one of the detectors at a unique angle
was unable to record $\gamma$ rays above 2~MeV, so this cancellation only works
below $E_{\gamma} = 2$~MeV.  Above that energy, uncertainties are higher due to
both angular correction uncertainties as well as larger uncertainties on
the absolute detector efficiency.  For all $\gamma$ lines below 2~MeV, the
measured cross sections from individual detectors $i$ are added together with appropriate coefficients
$\beta_{i}$ to cancel out anisotropic effects.  Above
$E_{\gamma} = 2$~MeV, the four detectors simply have their measured cross sections averaged, so
$\beta_{i} = 0.25$.
Our final cross section is then given as the weighted sum of that from individual
detectors after normalization to the iron measurement,
\begin{equation}
\sigma_{\gamma}\left(E_{n}\right) = \sum\limits_{i} \beta_{i} \sigma_{\gamma, i}\left(E_{n}\right)
\end{equation}

\subsection{Systematic Uncertainties}

Table~\ref{tbl:sysunc} summarizes the cross section measurement systematic uncertainties.
By design, large uncertainties associated with absolute detector efficiency and
absolute neutron flux are canceled by the iron normalization.  Thus, for
$E_{n} < 2~\mathrm{MeV}$, the systematic uncertainty is dominated by the
precision of the reference iron cross-section measurements.  We chose to treat
discrepancies in angular-correction measurements between different detectors 
from Ref.~\cite{Dietz:56festuff} as an angular uncertainty, making the
iron angular distribution a leading uncertainty term.

Angular-distribution uncertainties for xenon were estimated by assuming that the xenon
anisotropy should be similar to that of iron~\cite{Dietz:56festuff} or 
lead~\cite{Guiseppe:2008qj}, and using the largest anisotropy factors as a
conservative estimate.  Variations in measured xenon cross sections between 
individual HPGe detectors at different angles
are consistent with or smaller than the 18\% deviations found in other
measurements, so this uncertainty is appropriate.  For $E_{\gamma} < 2~\mathrm{MeV}$, 
this anisotropy is largely
canceled as described in Sec.~\ref{sec:absxs}, giving a smaller uncertainty.

\begin{table}[htb]
\centering
\caption[Systematic uncertainties due to \otsx measurement]{Evaluated sytematic uncertainties
for cross section measurements. Xenon angular effects
dominate the systematic uncertainty for $E_{\gamma} > 2~\mathrm{MeV}$, where cancelation
was impossible.  These relative uncertainties are uncorrelated, and are added in quadrature.}
\begin{tabular}{l c} 
\hline\hline
Source & Uncertainty \\
\hline
reference $\sigma_{\mathrm{Fe}}$ & 6.7\% \\
$\sigma_{\mathrm{Fe}}$ angular correction & 7.9\% \\
iron thickness & 2\% \\
iron/xenon efficiency difference & 4\% \\
$\gamma$ ray attenuation & 3\% \\
HPGe detector $\epsilon(\gamma)$  $\left(E_{\gamma} < 2~\mathrm{MeV}\right)$ & 3\%\\
HPGe detector $\epsilon(\gamma)$  $\left(E_{\gamma} > 2~\mathrm{MeV}\right)$ & 6\% \\
xenon thickness & 3.3\% \\
xenon angular effects $\left(E_{\gamma} < 2~\mathrm{MeV}\right)$ & 3\% \\
xenon angular effects $\left(E_{\gamma} > 2~\mathrm{MeV}\right)$ & 18\% \\
\hline
\end{tabular}
\label{tbl:sysunc}
\end{table}

Attenuation corrections were calculated using \mbox{MCNPX}~\cite{mcnpx}, and the
corrections are conservatively treated also as an uncertainty.  The geometric
efficiency differences between xenon and iron were computed with a simple custom
Monte Carlo simulation, and are kept small by the use of multiple detectors at different angles.
Detector $\gamma$-efficiency uncertainty is based on the efficiency fit to the
$^{152}$Eu data.  An additional
uncertainty is added above 2~MeV 
to account for deviations in the efficiency curve extrapolation~\cite{Blank:2014tka, MacMullin2013, Fotiades:2013ama}.
Xenon-thickness uncertainty comes from temperature uncertainty, changing xenon
pressure, and a slight ballooning of the kapton windows at the ends of the target vessel.

\subsection{\label{sec:metastable}Metastable state}

\begin{figure}[htb]
    \begin{center}
    \includegraphics[keepaspectratio=true, width=\columnwidth]{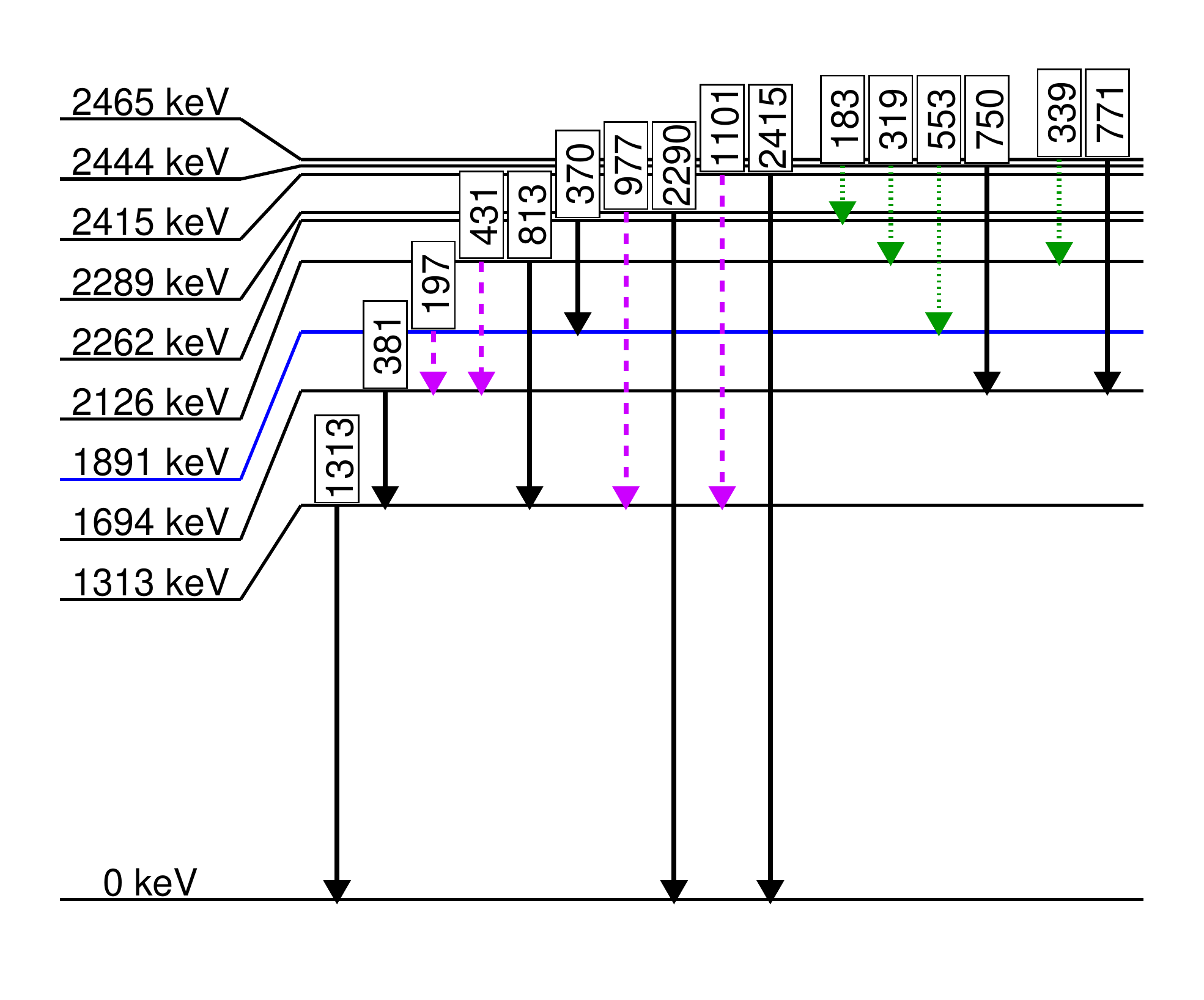}
    \end{center}
    \caption[136Xe Levels]
    {Diagram of levels up to 2500~keV, and their transitions, for \otsx.  
    Level scheme is taken from Ref.~\cite{SONZOGNI2002837}.  
    Energies are in keV.
    Transitions with solid lines have cross sections determined in this work.  Dashed lines indicate
    that the transitions were observed, but could not have cross sections precisely determined due to
    an interfering background line, low statistics, or other issue.  
    Dotted lines indicate that the
    $\gamma$ yield from those transitions was too weak to be clearly identified.
    The 1891~keV level is notable for its long half life of nearly 3~$\mu$s.
    (color online)}
     \label{fig:xelevels}
\end{figure}

\begin{figure}[htb]
    \begin{center}
    \includegraphics[clip, keepaspectratio=true, width=\columnwidth, trim={0.1cm 0.05cm 1.8cm 0.9cm}]{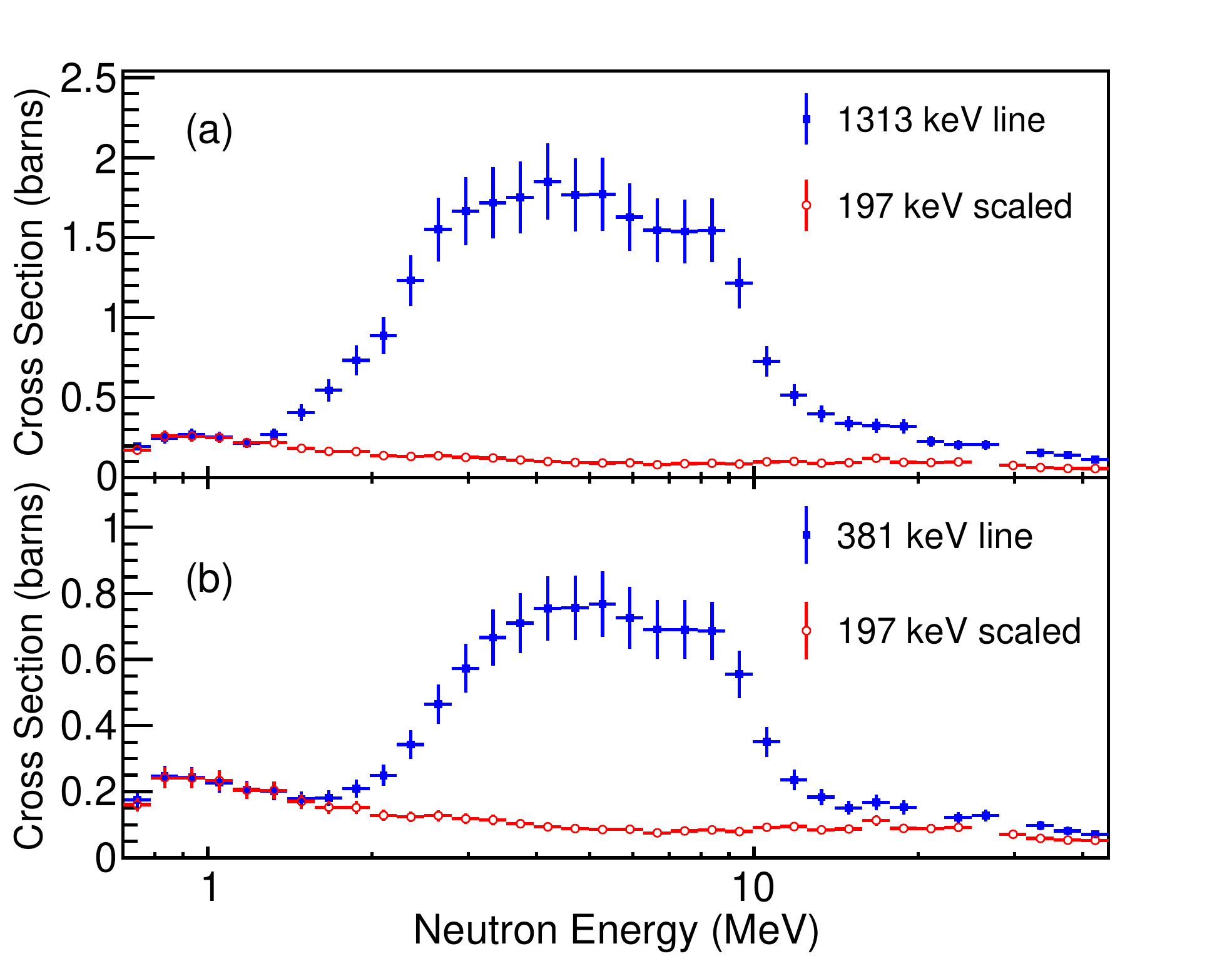}
    \end{center}
    \caption[metastable state subtraction]
    {Unsubtracted cross sections for the 1313~keV (a) and 381~keV (b) lines, along with a
    scaled cross section for the 197~keV line.
     The agreement of the cross section shape with time-of-flight corresponding to energies below $E_{n} = 1.3$~MeV
     suggests that the observed events below threshold are due entirely to the
     meta-stable state at 197~keV.  The final cross section evaluation (Fig.~\ref{fig:doublennp}) has
     the scaled 197~keV cross section subtracted off, leaving only prompt
     $\gamma$ emission.  Error bars here include 
     both statistical and systematic effects.
    (color online)}
     \label{fig:subtraction}
\end{figure}

Figure~\ref{fig:xelevels} shows the level diagram for \otsx, with levels up to
2500~keV.  Notably, the 1891~keV level is much longer-lived than the others, with
a nearly 3~$\mu$s half-life~\cite{SONZOGNI2002837}.  This meta-stable state interferes with the
time-of-flight measurement of $E_{n}$.  As a result, the 197~keV $\gamma$-emission
cross section cannot be properly measured, and there is a delayed component to
the 1313~keV and 381~keV lines that interferes.  We subtract
off the delayed component of these lines (which has the same time-of-flight distribution as
the 197~keV line) to leave only prompt $\gamma$ emission.  Figure~\ref{fig:subtraction}
shows this subtraction for both these lines.
While the full cross section for this delayed emission cannot be precisely
determined, we estimate from our measurements that 28\% (12\%) of emission is delayed for the 381~keV (1313~keV) line.

\section{\label{sec:inelscat_meas}Inelastic Scattering Measurements}

\begin{figure}[htb]
    \begin{center}
    \includegraphics[clip, keepaspectratio=true, width=\columnwidth, trim={0.1cm 0.00cm 1.8cm 0.9cm}]{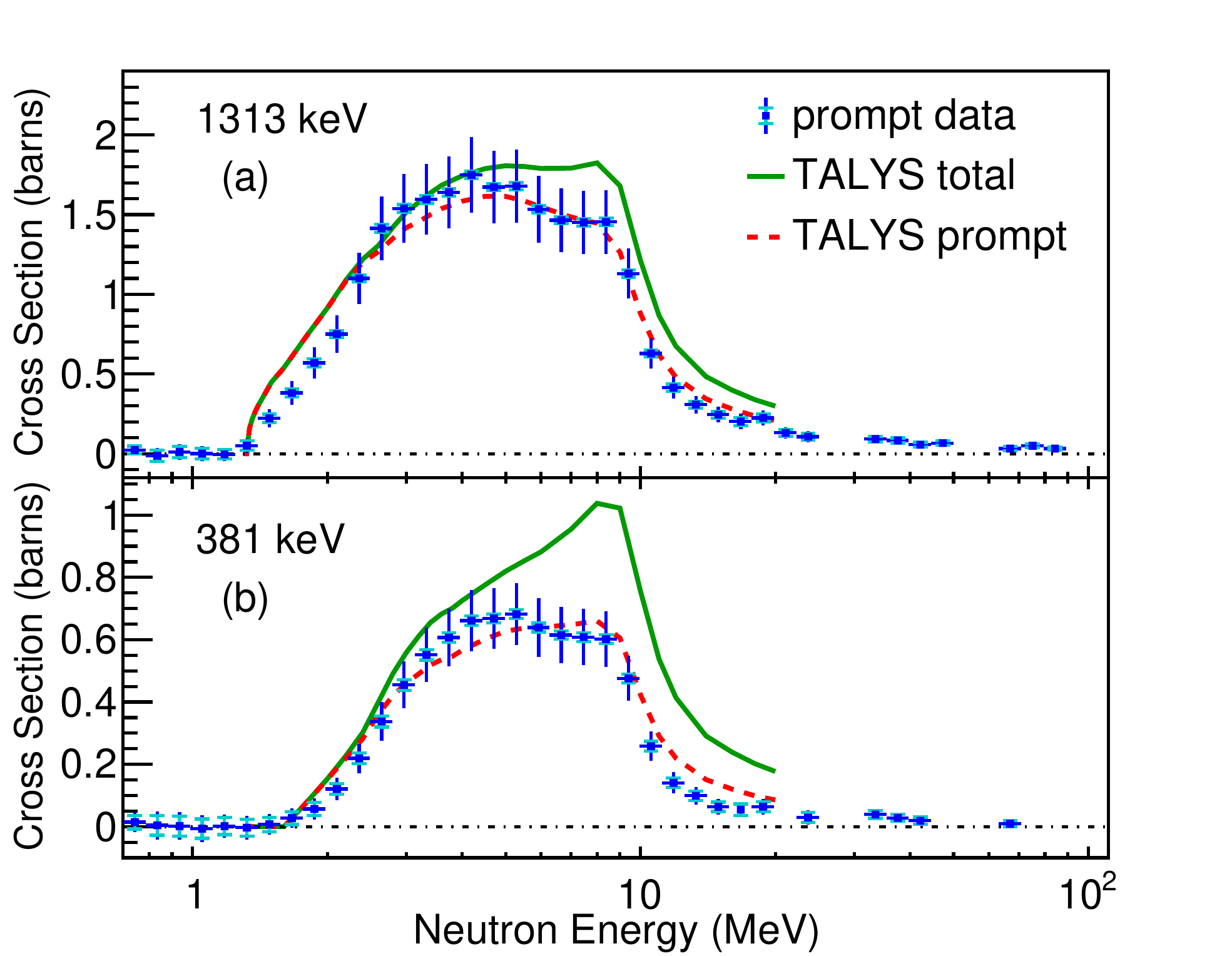}
    \end{center}
    \caption[n35n]
    {Measured cross sections for the two dominant $\gamma$-ray production lines,
    1313~keV (a) and 381~keV (b), from $(n,n')$.  Measured prompt gamma ray production
    cross sections are indicated with blue square markers and error bars, 
    with delayed production subtracted out as described in
    Sec.~\ref{sec:metastable}.  Blue (cyan) error bars
    here, and in subsequent figures, indicate combined statistical and systematic errors
    (statistical errors alone).
    The TALYS~1.8 prediction for total (prompt) gamma ray production
    is indicated with a solid green line (dashed red line). 
    (color online)}
     \label{fig:doublennp}
\end{figure}

\begin{figure*}[htb]
    \begin{center}
    \includegraphics[clip, keepaspectratio=true, width=\textwidth, trim={0.1cm 0.00cm 0.0cm 0.5cm}]{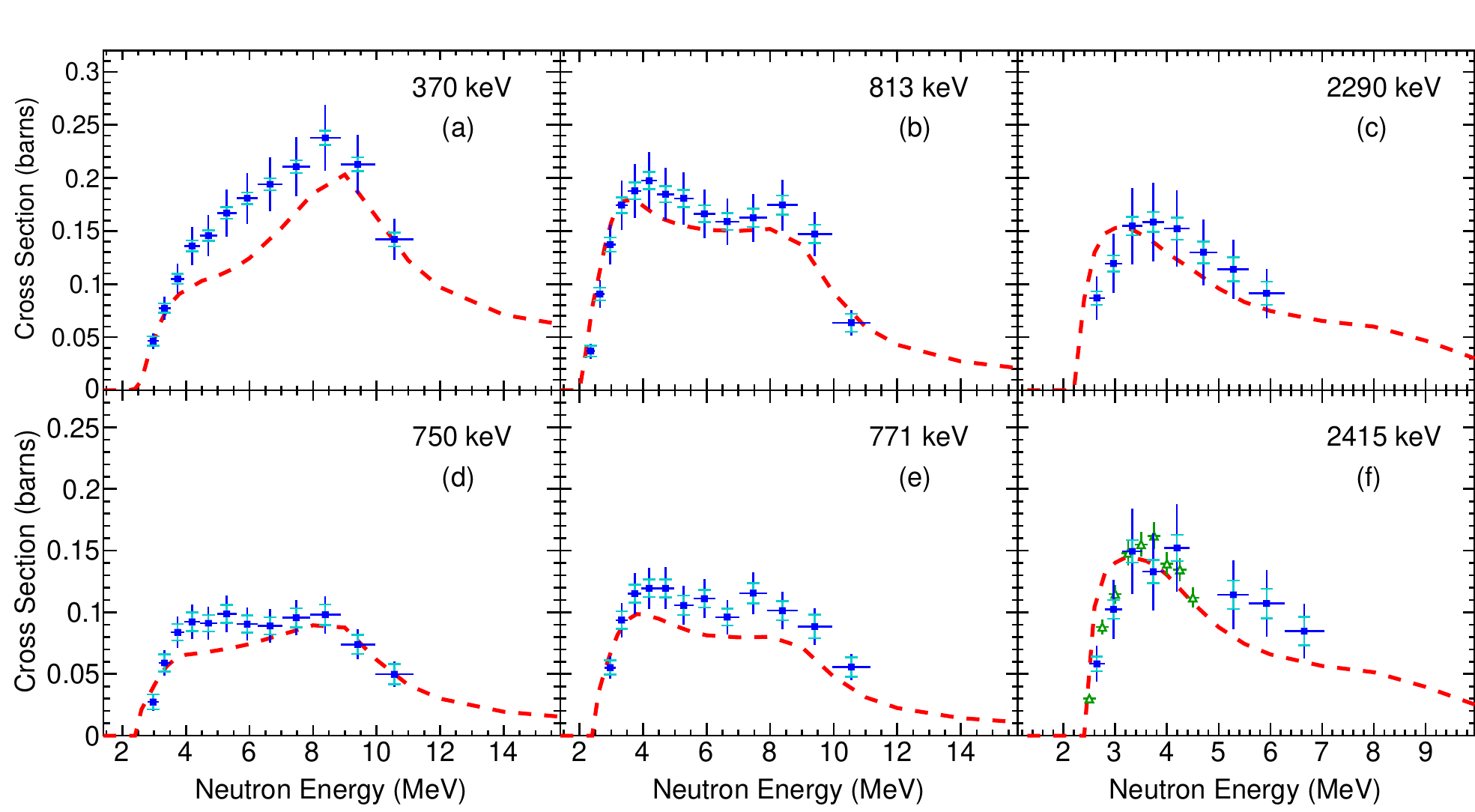}
    \end{center}
    \caption[nnp]
    {Measured cross sections for $\gamma$-ray production from several $(n, n')$
    inelastic scattering interactions.  Measured values are indicated with 
    blue square markers and error bars, while TALYS~1.8 predictions
    are red dashed lines.  
    The 2415~keV measurement (f) additionally is compared to recent results from
    Peters~\textit{et al.}~\cite{Peters:2017xtp}, indicated by open green
    triangles.
    (color online)}
     \label{fig:sixplot_nnp}
\end{figure*}

\begin{figure}[htb]
    \begin{center}
    \includegraphics[clip, keepaspectratio=true, width=\columnwidth, trim={0.1cm 0.00cm 1.8cm 0.9cm}]{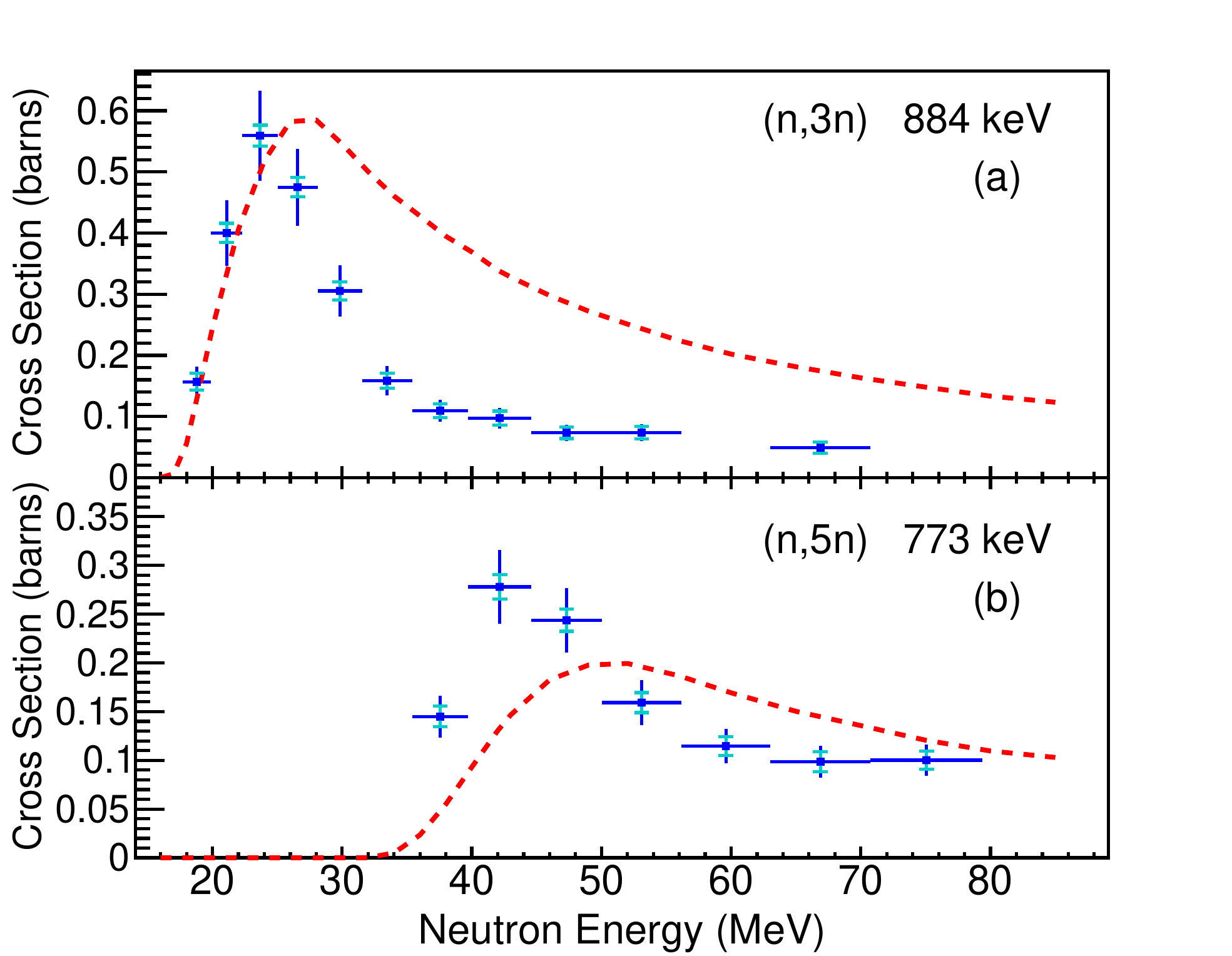}
    \end{center}
    \caption[n35n]
    {Measured cross sections for $\gamma$-ray production from $(n, 3n)$ and
    $(n, 5n)$ interactions.  Measured values are indicated with 
    blue square markers and error bars, while TALYS~1.8 predictions
    are red dashed lines.
    (color online)}
     \label{fig:doublen35n}
\end{figure}

Cross sections could only be determined for $\gamma$ lines with sufficient yield 
to be fit successfully in all available detectors, 
and only if a $\gamma$ ray of similar intensity and energy
is not present in background.  After these selections,  we are left with 
eight lines from $(n,n')$ interactions.  These measured cross sections are shown
in Fig.~\ref{fig:doublennp} and~Fig.~\ref{fig:sixplot_nnp}, along with estimates
from TALYS 1.8~\cite{talys}.  Several other $\gamma$
lines were observed, with most of them identified as being due to $(n,xn)$
interactions for $x = 2, 3, 4, 5$.  Most of these either had low statistics or
an interfering background, and so a full cross section determination was not possible.  
The cross section for one line each from
$(n, 3n)$ and $(n,5n)$ could be determined, and are shown in 
Fig.~\ref{fig:doublen35n}.  In these figures, bins in $E_{n}$ with no marker do not necessarily
indicate zero cross section, but may be bins where, due to low statistics, 
the peak fit for data from one or more detectors failed.
Lines at 1640, 1690, and 2009~keV appeared for $E_{n} < 5$~MeV and were 
identified as likely being from $^{136}\mathrm{Xe}(n,n')$,
though were not measurable due to interfering nearby lines or low statistics.  These lines
do not correspond to any obvious known
transition in the ENSDF database~\cite{10.1007/978-3-642-58113-7_227}.  
Another unknown line at 1355~keV was seen, though it only appeared
with $E_{n} > 10$~MeV, and is likely an unknown line from some daughter nucleus
of a neutron-induced reaction.

\section{\label{sec:limits}Limits on new $\gamma$ lines relevant to \znbb}

\begin{figure}[htb]
    \begin{center}
    \includegraphics[clip, keepaspectratio=true, width=\columnwidth, trim={0.05cm 0.05cm 1.0cm 0.7cm}]{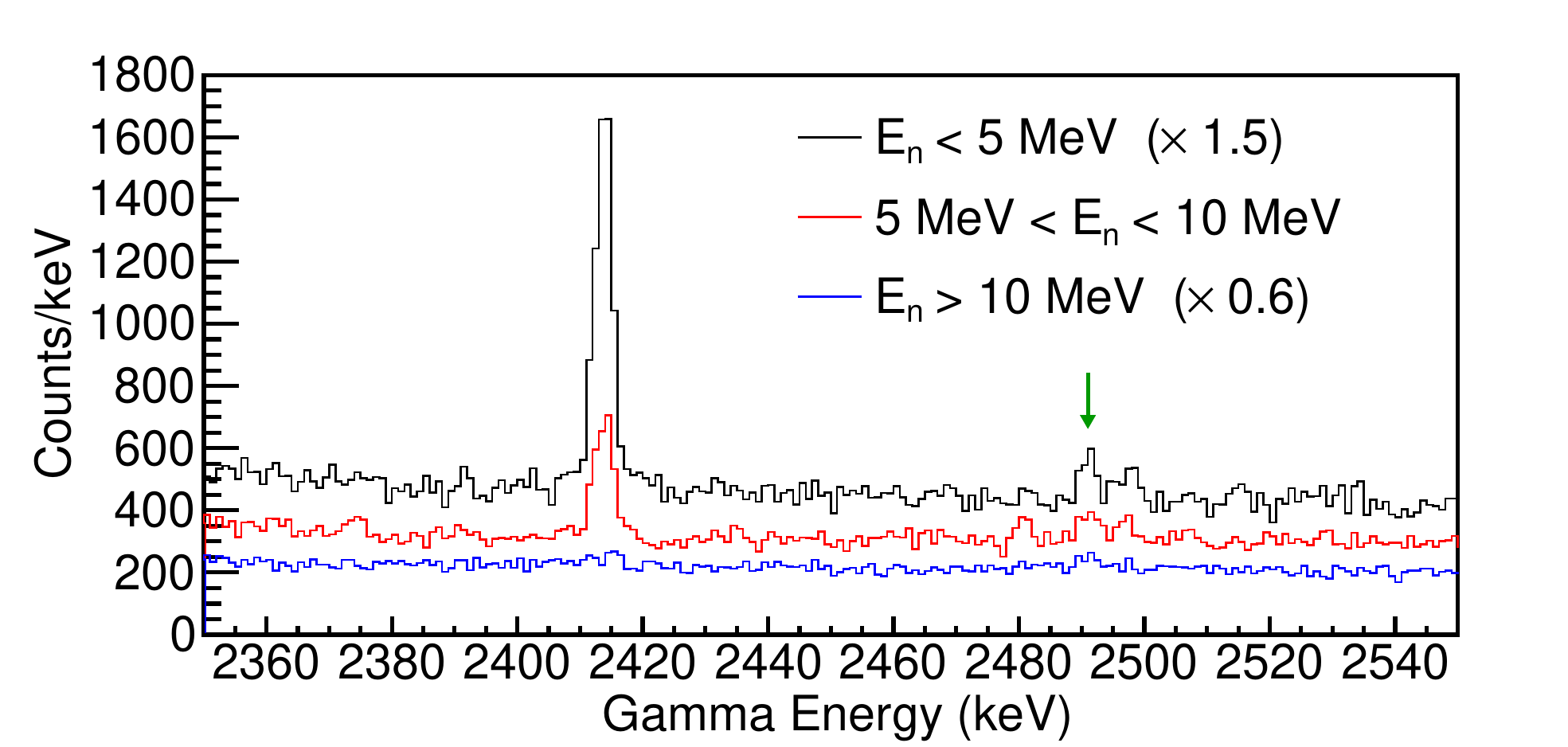}
    \end{center}
    \caption[roi search]
    {Sum of detected counts in four HPGe detectors for $E_{\gamma}$ in the general
    region of interest for \otsx \znbb experiments.  The spectra have been separated
    into three neutron energy ranges via scaling as they were in Fig.~\ref{fig:detPexamples}(a).
    A peak near 2491~keV (possibly due to inelastic scattering on the 
    germanium detectors) is indicated with an arrow.
    (color online)}
     \label{fig:roisearch}
\end{figure}

The discovery of a new $\gamma$ line produced from neutron interactions on \otsx
could be an important consideration for next-generation \znbb experiments.  
As such, we have searched for $\gamma$ peaks
between 2350 and 2550~keV around the \otsx \znbb Q-value of 2457.8~keV~\cite{Redshaw:2007un}.
The count spectra from the four HPGe detectors summed is shown in Fig.~\ref{fig:roisearch}.
For the calculation of cross-section limits, we use only data from one detector
(systematic uncertainties are larger than statistical ones here).  
Limits are based on a sliding 5~keV $E_{\gamma}$ window, in
which counts above background are considered with Poisson statistics to
determine a 90\% upper limit on peak intensity.  The rest of the cross-section
determination and systematic uncertainties are as described in Sec.~\ref{sec:data_ana}.

There is a known line at 2415~keV from a transition to the ground state of \otsx.  This
is separated in energy from the Q-value by 1.8\%.  Our measurements of this line
are shown in Fig.~\ref{fig:sixplot_nnp}(f).  
The rest of the
spectrum is relatively flat, though there is a small peak  
near 2491~keV, 1.4\% away from the Q-value.  This peak is seen in all three $E_{n}$
regions, and we determined cross section upper limits on this peak
to be 13, 48, and 37~mb in the $E_{n} < 5$~MeV, $5~$MeV$<E_{n} < 10$~MeV,
and $E_{n} > 10~$MeV ranges, respectively.  Note that these limits are for
cross sections averaged
over the neutron energy range, weighted by the GEANIE flux.    
The source of this peak is unknown, but its energy matches a known transition in
$^{72}\mathrm{Ge}$, and a search for this peak in our non-xenon data was inconclusive.  
No other new peaks of comparable magnitude were
found near the \otsx Q-value, so these cross-section limits hold for any new
peaks in this $\gamma$ energy range.

\section{\label{sec:technique}Novel Calibration Technique}

Neutron-induced excitation to the 1313~keV level is the primary $\gamma$-producing
interaction observed for fast neutrons below~$\sim$2.5~MeV in energy.  Thus, neutron inelastic
scattering could potentially be a source of mono-energetic 1313~keV $\gamma$ rays for
calibration of a future \znbb detector.  A great strength of proposed experiments,
such as nEXO~\cite{Kharusi:2018eqi}, is the pure, monolithic, self-shielding, active
detector volume.  Without radioisotopes being introduced into the detector, very few
$\gamma$ rays will reach the central part of the xenon volume.  This is excellent
for background reduction, but potentially problematic for calibration.

It may be possible to use DD (deuterium-deuterium) generator located outside
the xenon volume to deliver 2.5~MeV neutrons deep into the detector, exciting \otsx
nuclei and producing 1313~keV $\gamma$ rays.  The neutrons from the
DD generator will scatter in the xenon both elastically and inelastically.  Elastic
scatters will tend to deposit small amounts of energy in nuclear recoils, but most
neutrons will inelastically scatter or leave the xenon volume before 
slowing down to an energy below the threshold for inelastic scattering.
Using a Geant4~\cite{Agostinelli:2002hh,Allison:2006ve} 
simulation with liquid xenon enriched to 
90\% \otsx, we found the 
effective interaction length for DD generator neutrons to inelastically scatter
to be 23~cm.  This compares very favorably to $\gamma$ rays, which have an interaction length
below 10~cm.  Thus, this could be a useful alternative to external $\gamma$ sources
for calibrating detector
response for interactions near the detector center.  While calibration using
neutron inelastic scatters has been proposed for nuclear recoil 
measurements in liquid noble gas detectors~\cite{1748-0221-9-10-P10017}, we
believe this would be a new technique for detectors designed for electron
recoil measurements.

There would be some challenges in implementing this calibration method.  The
elastic scattering would deposit additional energy in the detector beyond the
mono-energetic 1313~keV line.  However, these depositions will produce only small
amounts of scintillation and ionization (reduced further by quenching), and the position-dependence of this 
could be corrected for via
simulations which do not depend on the detector response, only the neutron
propagation physics and quenching factors.  The generator would need to be
brought near the xenon volume in such a way as to avoid negatively impacting radiopurity
during normal running.  Neutron captures could likely be excluded from calibration data based on
interaction timing, but possible activation or neutron damage to electronics, while
unlikely with neutrons below 3~MeV, would need to be considered.  Finally, the
costs and benefits of this method must be weighed against other techniques for
calibrating deep inside detectors, including external $\gamma$-ray calibration and
dissolving a radioisotope in the xenon volume~\cite{Aprile:2016pmc}.

\section{\label{sec:conclusions}Discussion and Summary}
The $(n,n')$ cross sections measured here are similar to estimates from TALYS simulations.
The $(n,xn)$ cross sections are not well estimated by TALYS, so this measurement may suggest possible
improvements to reaction modeling.  All TALYS predictions presented in this work 
were obtained with the TALYS~1.8 default settings.  We have not attempted to test 
the dependence of the predictions on different models or involved quantities.
These measurements, along with
newly observed $\gamma$ lines, may also be used to refine future
nuclear evaluations and improve simulations.
The measured cross section for the 2415~keV 
line matches well with recent measurements using $E_{n} = 2.5 - 4.5$~MeV made by Peters~\textit{et al.}~\cite{Peters:2017xtp}.
The $E_{n}$ dependence of the cross sections for the 1313~keV and 884~keV lines
agree within uncertainties with unnormalized measurements from Fotiades~\textit{et al.}~\cite{Fotiades:2007rn}.

We have searched for new $\gamma$ lines near the \otsx Q-value, and set limits
on their possible neutron interaction cross sections for a wide range of $E_{n}$.  
Any lines which may interfere with \znbb measurement
have considerably smaller cross sections than the known line at 2415~keV.  
Thus, potential fast-neutron-induced backgrounds
to \znbb have been well-catalogued, and are unlikely to impact future experiments.

The calibration technique proposed here for large liquid noble gas detectors
 using neutron inelastic scattering
to provide a mono-energetic $\gamma$-ray could be valuable for future experiments.
The costs and challenges of this technique will need to be weighed against
existing techniques to determine viability, but, for relevant experiments, it should
be worth considering.

\vspace{3mm}

\section{Acknowledgments}

This work was funded by US Department of Energy Grant No. DE-SC0012191.
This work was performed under the auspices of the U.S. Department of Energy
(DOE) under Contract No. DE-AC52-06NA25396. This work has benefitted from use of
the LANSCE accelerator facility supported under DOE Contract No.
DE-AC52-06NA25396.  
We thank Vladimir Belov for his help studying neutron inelastic scattering as a calibration 
technique.

\bibliographystyle{./apsrev4-1}
\bibliography{geanie_paper}

\end{document}